\documentclass{ws-procs9x6}
\usepackage{epsfig}

\begin{document}

\title{Physics of Heavy Flavour at CDF}

\author{STEFANO TORRE\footnote{\uppercase{O}n behalf of the \uppercase{CDF} \uppercase{C}ollaboration.}}

\address{Univerit\`a degli Studi di Siena\\
Dipartimento di Fisica\\
Via Roma 56 - 53100 Siena - Italy\\
and\\
Istituto Nazionale di Fisica Nucleare sez. di Pisa \\
Largo B. Pontecorvo 3 - 56100 Pisa - Italy \\
E-mail: stefano.torre@pi.infn.it}

\maketitle

\abstracts{  Results   on  physics  of   heavy  flavour  at   CDF  are
reported. Selected measurements of Branching Ratios and CP asymmetry in
$B^0$ and $B_s^0$, lifetime difference of $B^0_s$ CP eigenstates and a
precise measurement of the $B_c$ mass are presented.}

\section{Introduction}
The  upgraded  Collider  Detector  at Fermilab  (CDF~II)\cite{cdf}  has
collected  around   $800\,\rm{pb^{-1}}$  between  February   2002  and
February 2005 during  the Tevatron Run II at  Fermilab. 
At $p\bar{p}$ colliders a large amount of $b$ and $c$ mesons and baryons
 are produces within a background of hadronic particles. However the
presence of heavy flavour particle's decays can be detected by the presence 
of a displaced secondary vertex, because of these particles have long 
decay length (O($100\mu m$)). The issue is to be able to extract this 
information at trigger level. For this purpose CDF uses the 
 Silicon  Vertex Trigger  (SVT)\cite{svt} that reconstruct online the 
tracks providing the informations needed for the trigger decision. In this 
way CDF is able to efficiently select events in which the heavy meson decays 
in either leptonic or fully hadronic modes. The collected data samples 
allow to perform a wide range of measurements, 
from the observation
of rare decays to lifetime  measurement, through BR and $CP$ asymmetry
measurements. In the following we concentrate on some selected topics.

\section{Branching ratios and CP asymmetries measurements}
Fully hadronic $b$ meson decays  are very useful to understand the $b$
sector  of the CKM  matrix. CDF  is providing  interesting measurement
both on two body charmless and on pure penguins decays.

\subsection{Two body charmless decays ($B\rightarrow h^+ h'^-$)}
\begin{figure}[ht]
\centerline{\epsfxsize=4cm\epsfbox{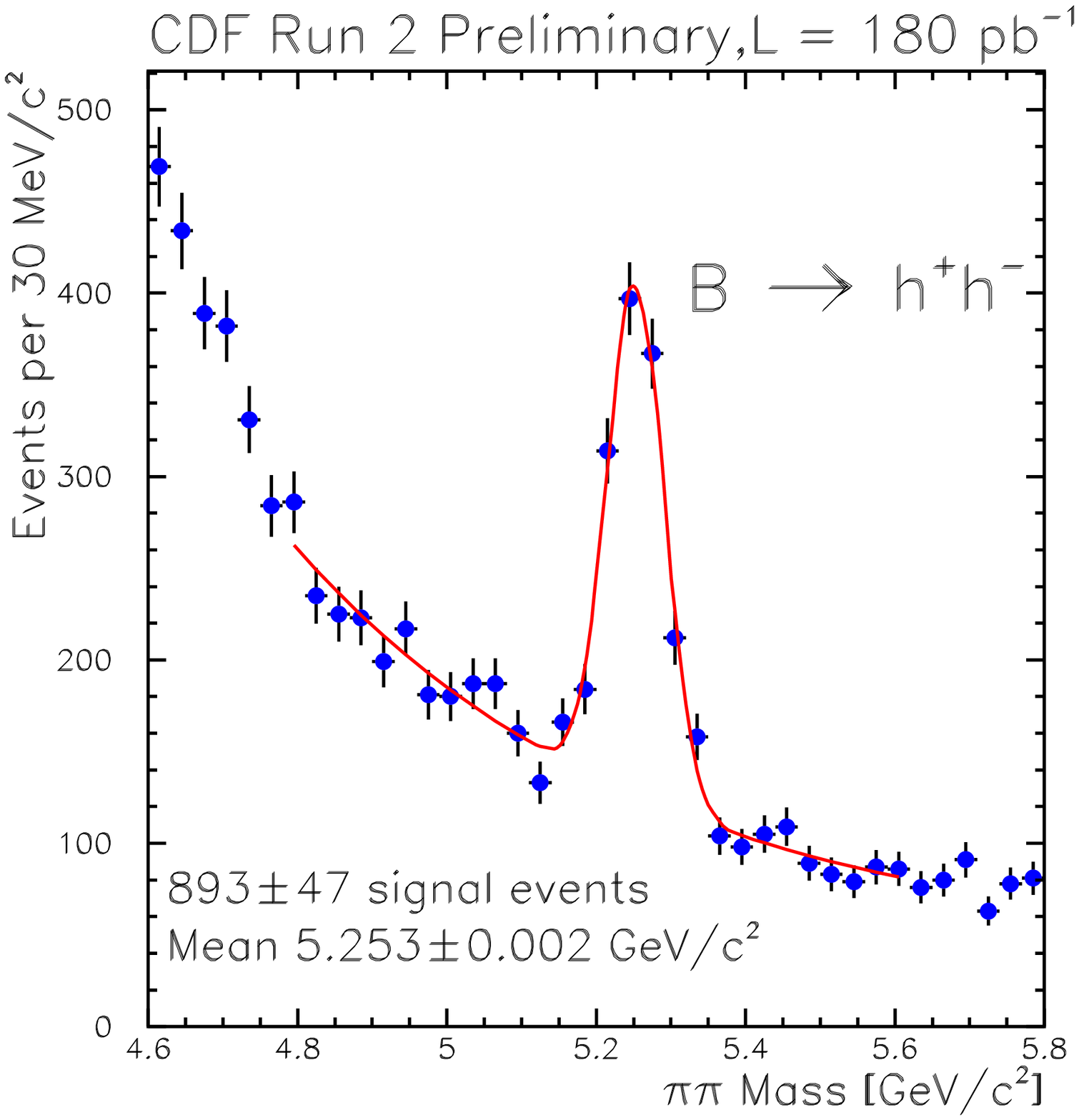}\epsfxsize=4cm\epsfbox{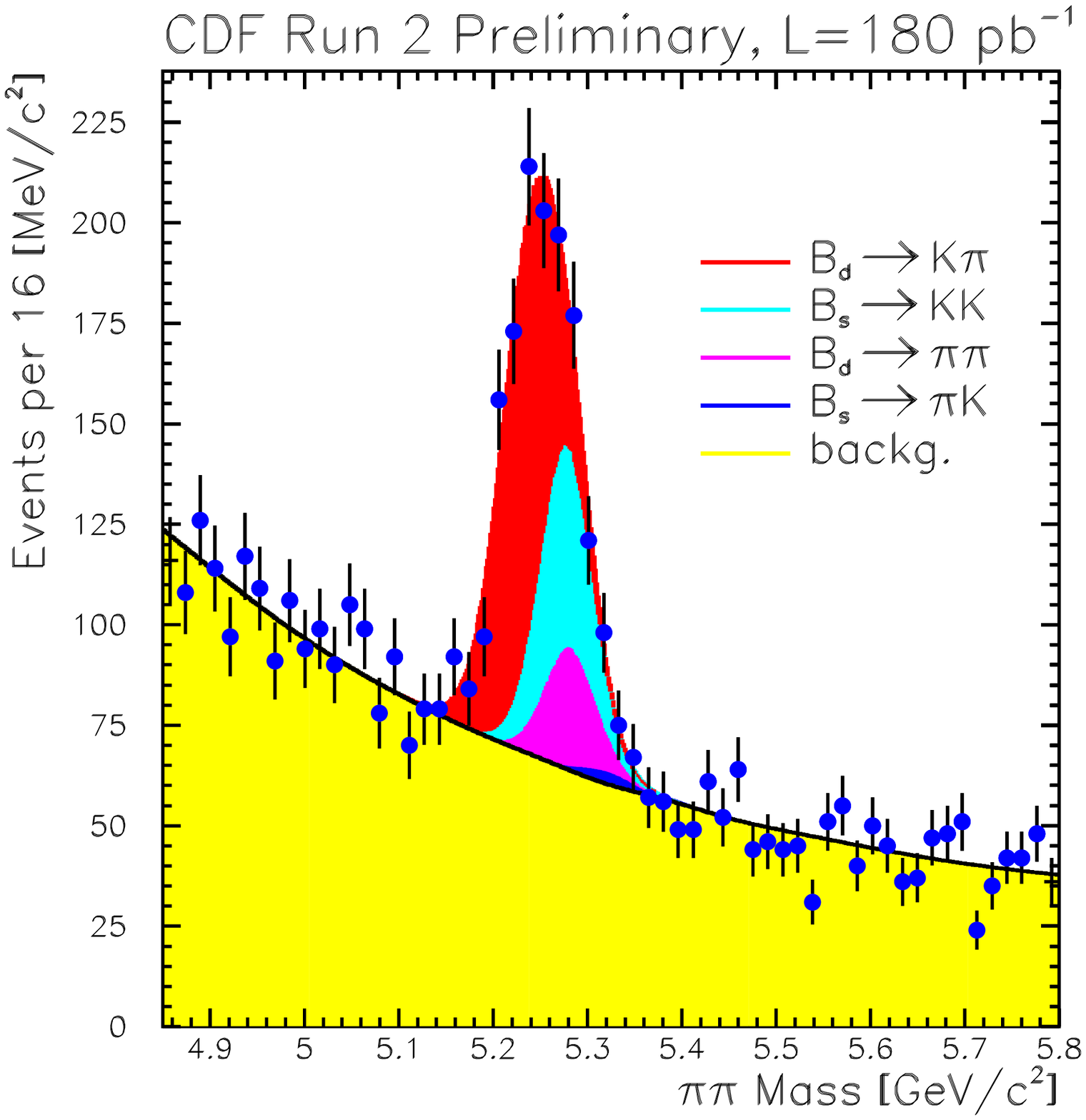} }   
\caption{\label{Bhh} The plot on the left shows yield of $B\rightarrow
h^+ h'^-$ events at CDF. The signal is given by several decays. By
an  unbinned  likelihood  fit   the  different  contributions  can  be
extracted and  the plot  on the right  shows, in different  color, the
different contributes.}
\end{figure}

These decays are  the ones in which a $B^0$,  $B_s^0$ meson goes into
charged Pions and  Kaons. The theoretical prediction on  their BR and CP
asymmetries 
are strongly affected by uncertainties on
hadronic contributes.  These unknowns  can be removed by combining the
informations obtained in the different modes\cite{bhhteo,bvvteo2}.

The  mass resolution  at CDF  is not  enough to  directly  observe the
different signals, but  their yields can be extracted  via an unbinned
likelihood  fit   that  exploit  both  kinematic   and  energy  loss
information.   The  overall  yield  is  shown  in  the  left  plot  in
Fig.\ref{Bhh}.  The result of the fit  is shown in the  right plot of
Fig.\ref{Bhh}.    In   $180\,\rm{pb}^{-1}$   we    observe   $509\quad
B^0\rightarrow  K^+\pi^-$,  $134\quad  B^0\rightarrow \pi^+\pi^-$  and
$232\quad   B_s^0\rightarrow   K^+K^-$.    We  measured   the   ratio:
$\frac{f_s\mathcal{B}(B_s^0\rightarrow
K^+K^-)}{f_d\mathcal{B}(B^0\rightarrow  K^+\pi^-)}=0.50  \pm0.08(stat)
\pm 0.07(syst)$.   We set also the  limits on the BRs  of rare $B_s^0$
decays            as:           $\frac{f_s\mathcal{B}(B_s^0\rightarrow
K^+\pi^-)}{f_d\mathcal{B}(B^0\rightarrow      K^+\pi^-)}<0.11$     and
$\frac{\mathcal{B}(B_s^0\rightarrow
\pi^+\pi^-)}{\mathcal{B}(B_s^0\rightarrow K^+K^-)}<0.10$ both at $90\%
C.L.$.   We  also measure  the  CP  asymmetry  in the  $B^0\rightarrow
K^\pm\pi^mp$ decay and we obtain $-0.04\pm 0.08(stat)\pm 0.01(syst)$.

\subsection{Pure penguin decays} 
\begin{figure}[ht]
\centerline{\epsfig{file=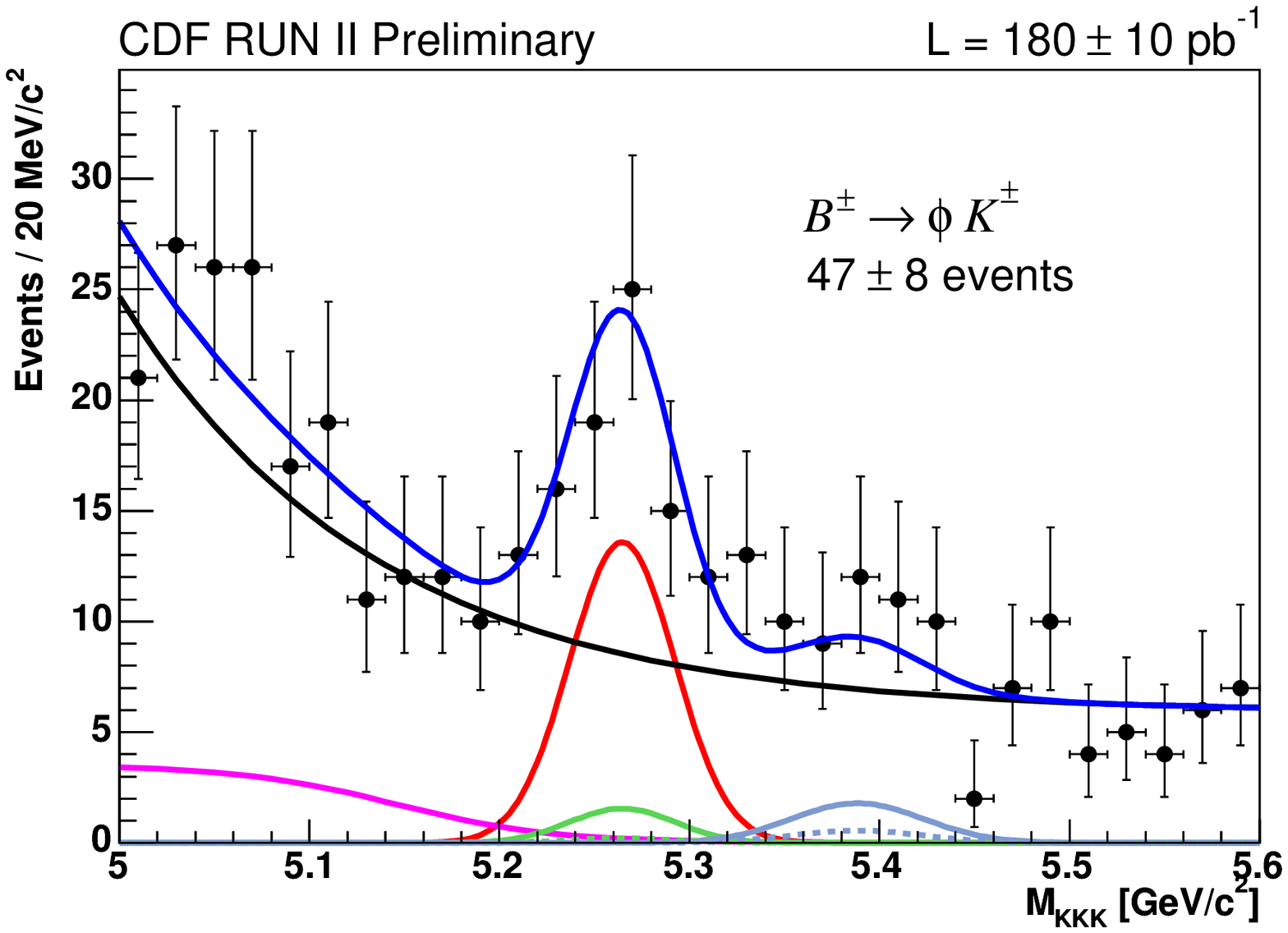,height=4cm}\epsfig{file=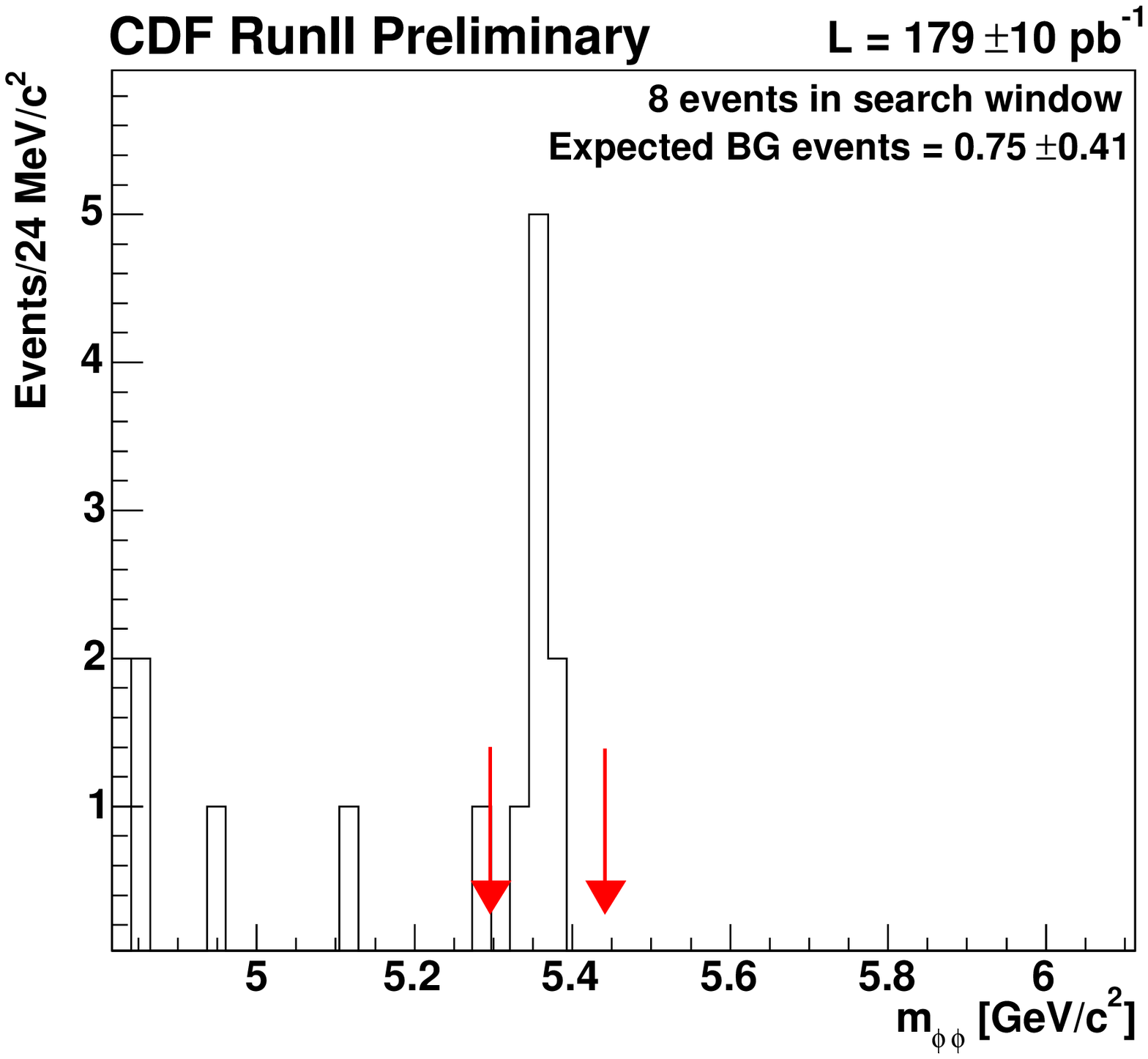,height=4cm}}   
\caption{\label{BVV}  The  plot  on  the  left  shows  the  signal  of
$B^\pm\rightarrow\phi  K^\pm$  events. The  blue  line  is the  fitted
distribution of signal and background events. The red line is the only
signal contribute to the fit. The other lines are different background
contributes.   The  plot   on  the   right  shows   the   evidence  of
$B_s^0\rightarrow\phi\phi$ events.}
\end{figure}
B  meson  decays involving  $b\rightarrow  s\bar{s}s$ transitions  can
provide  evidences   of  deviation  from   the  SM\cite{bvvteo1}.   In
particular direct CP asymmetry of $B^\pm\rightarrow\phi K^\pm$ mode is
expected   to  be   of   the   order  of   few   percent  within   the
SM\cite{bvvteo2}.  Left  plot  in   Fig.   \ref{BVV}  shows  the  mass
distribution containing the signal of this decays obtained at CDF. The
number of signal  events has been extracted from  the background using
an unbinned likelihood fit  on kinematic the particle's energy losses.
We     measured     $\mathcal{B}(B^\pm\rightarrow\phi    K^\pm)=7.6\pm
1.3(stat)\pm        0.6(syst)        \times        10^{-6}$        and
$a_{CP}(B^\pm\rightarrow\phi
K^\pm)=-0.07\pm0.17(stat)^{+0.03}_{-0.02}(syst)$.

The   same  $b$   transition   occurs  in   $B_s^0\rightarrow\phi\phi$
decay. First  evidence of  this decay  has been found  at CDF  and the
right plot in  Fig.  \ref{BVV} shows the observer  signal.  We measure
$\mathcal{B}(B^\pm\rightarrow\phi\phi)=1.4^{+0.6}_{-0.5}(stat)\pm
0.2(syst)\pm0.5(BR)\times 10^{-6}$ where the  last error come from the
BR  of  the  $B_s^0\rightarrow  J/\psi\phi$  that  has  been  used  as
normalization  mode.   Details  of  these  analysis can  be  found  in
Ref.\cite{bvvpaper}. 

\section{Measurement of decay width  difference of $B_s^0$ CP eigenstates}
\begin{figure}[ht]
\centerline{\epsfig{file=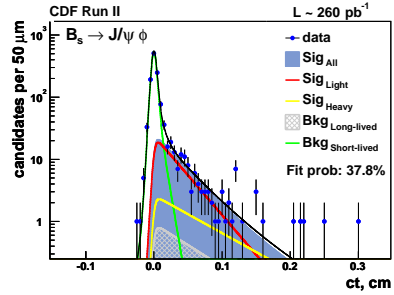,height=4cm}}   
\caption{\label{deltags} Projection of the unbinned maximum likelihood
fit on  the lifetime distribution. The red/yellow  line correspond to
the light/heavy mass eigenstate.}
\end{figure}
In $B_s^0\rightarrow J/\psi\phi$ two  vector particles are produced from
the decay of a pseudo scalar  one. It is indeed possible to distinguish
the two  CP eigenstates by  the relative angular distribution  of the
decay's products. 

Actually three linear amplitudes are possible corresponding to angular
momenta 0,1 and 2. The  even/odd values of angular momentum occur for
CP even/odd  eigenstate. We simultaneously  fit the lifetime  and the
linear  amplitudes.   The  plot   in  Fig.   \ref{deltags}  shows  the
projections of the fit result on  the lifetime. The yellow line is the
lifetime of  the heavy mass eigenstate  while the lifetime of  the light
one      is      reported      in     red.       We      found
$\frac{\Delta\Gamma_s}{\Gamma_s}=65^{+25}_{-33}\%$\cite{dgspaper}.

\section{Spectroscopy}
\begin{figure}[ht]
\centerline{\epsfig{file=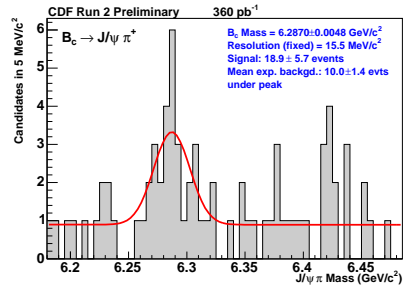,height=4cm}}   
\caption{\label{bc}Mass  distribution of events  in the  search window
for $B_c \rightarrow  J/\psi \pi$. The red line is  the fit applied to
data to measure the mass.}
\end{figure}
In  the field of  spectroscopy the  measurement of  the $B_c$  mass is
important  to  validate  the  theoretical  models  that  predict  this
quantity \cite{bcteo}. This  meson has
been observed  at D0\cite{bcd0} and  CDF\cite{bccdf}  in semileptonic
modes.   Because in  this  kind of  decays  the events  are not  fully
reconstructed the achieved mass  resolution was not enough to constrain
the  theoretical  predictions.   We  look  for evidence  of  the  fully
reconstructed  $B_c \rightarrow J/\psi  \pi$ decay  in the  mass range
between  $5.6$  and  $7.2\,\rm{GeV/c^2}$  corresponding  to  $2\sigma$
window around the previously measured value.

We optimize the  selections following a blind procedure.   The plot in
Fig. \ref{bc}  show the mass  distribution in the signal  region after
applying  the optimized  cut. From  the fit  we measured  a  signal of
$18.9\pm 5.7$ events  over a background of $10.0\pm1.4$.  The value of
the mass we obtain is $6287.0\pm 4.8(stat)\pm 1.1(syst)\,\rm{MeV/c}^2$ where
the systematic  error is  mainly given by  the parametrization  of the
background.

\section{Conclusions}
We  have reported some  examples of  the wide  range of  heavy flavour
particles that can be detected at CDF and how their characteristics can
be    investigated.   This    analysis    are   still    statistically
limited.  However  the systematics  are  well  under  control and  the
results will be easily improved by increasing the data samples.

\end{document}